\documentclass[12pt]{article}
\usepackage[a4paper, margin=3cm]{geometry}
\usepackage[utf8]{inputenc}
\usepackage{preamble}

\title{\vspace{-2cm}\huge Are All Events Created At-Once in Relational Quantum Mechanics?}
\author{\orcidlink{0009-0002-1131-7559}Pablo Toussaint\thanks{Ludwig Maximilian University of Munich, E-Mail:\href{mailto:
toussaint.pablo@proton.me}{toussaint.pablo@proton.me}}}
\date{\normalsize{November 29, 2023}\vspace{-3ex}}

\begin{document}
\maketitle

\begin{abstract}\vspace{-1ex}

This paper discusses the possibility of temporal generation of events in relational quantum mechanics (RQM). It critically examines claims by Adlam and Rovelli that the events in RQM must have been created all-at-once in order to avoid a contradiction with the theory of relativity. The analysis demonstrates that not considering the set of events as absolute and observer-independent allows for their temporal generation. Furthermore, the paper establishes that even with the postulate of cross-perspective links, it remains possible to regard the set of events as non-absolute. Thus, events in relational quantum mechanics can be generated temporally and need not have been created all-at-once.


\end{abstract}

\section{Introduction}\label{sec: Introduction}

Quantum mechanics is incredibly successful at making predictions about the microscopic world. However, quantum mechanics as taught in textbooks is more of a recipe for making predictions than a physical theory, as a physical theory should make clear what it is talking about, that is, provide an ontology, and make clear how what it is talking about behaves, that is, provide dynamics \parencite{Maudlin2019PhilosophyPhysics}. To make quantum mechanics a complete physical theory, numerous ‘interpretations’ have been developed, one of which is relational quantum mechanics (RQM), which was introduced by \textcite{Rovelli1996RelationalQuantum} and is this paper’s subject.

The fundamental ontology of RQM consists of events or facts that occur in the interaction of two systems. These events correspond to measurements in standard quantum mechanics. However, the events of RQM can occur in the interaction of \emph{any} two systems. Classical systems or conscious ‘observers’ do not have a prominent role in RQM. For this reason, RQM is sometimes referred to as the Copenhagen interpretation \say{democratised} \parencite{sep-qm-relational}.

The relative aspect of RQM is that the events or the results of these \say{measurements} are valid only relative to the two interacting systems. The facts become approximately stable in the macroscopic world through decoherence effects, so that one can dispense with these labels in everyday life \parencite{Rovelli2021StableFacts}.
RQM thus describes relative facts that occur when two systems interact.\footnote{The role of systems in the Ontology of RQM is still unclear. For more on this, see \parencite{Rovelli2022InformationPhysical, Rovelli2023InformationPhysical, Morganti2022WhatOntology}.}

Since there are no intrinsic states in RQM, no state can evolve dynamically. However, RQM makes empirical predictions through transition amplitudes or transition probabilities between two facts that can potentially evolve over time.\\

As RQM is constantly evolving, this paper refers specifically to the version of RQM characterized by the following six postulates, that were introduced by \textcite{Pienaar2021QuintetQuandaries} and modified by \textcite{Rovelli2022RelationalQuantum, Rovelli2022InformationPhysical, Rovelli2023InformationPhysical}.

\begin{enumerate}
	\item \textbf{Relative facts:} Events, or facts, can happen relative to any physical system.
	\item \textbf{No hidden variables:} Unitary quantum mechanics is complete.
	\item \textbf{Relations are intrinsic:} The relation between any two systems $A$ and $B$  is independent of anything that happens outside these systems’ perspectives.
	\item \textbf{Cross-perspective links:} In a scenario where some observer Alice measures a variable $V$ of a system $S$, then provided that Alice does not undergo any interactions that destroy the information about $V$ stored in Alice’s physical variables, if Bob subsequently measures the physical variable representing Alice’s information about the variable $V$, then Bob’s measurement result will match Alice’s measurement result.
	\item \textbf{Measurement:} An interaction between two systems results in a correlation within the interactions between these two systems and a third one; that is, with respect to a third system $W$, the interaction between the two systems $S$ and $F$ is described by a unitary evolution that potentially entangles the quantum states of $S$ and $F$.
	\item \textbf{Internally consistent descriptions:} In a scenario where $F$ measures $S$, and $W$ also measures $S$ in the same basis, and $W$ then interacts with $F$ to \say{check the reading} of a pointer variable (i.e., by measuring $F$ in the appropriate \say{pointer basis}), the two values found are in agreement.\footnote{The postulates were taken unchanged from \textcite{Rovelli2023InformationPhysical}.}
\end{enumerate}

The postulate of cross-perspective links (CPLs) is formulated asymmetrically in time. Nevertheless, in the version of the paper introducing CPLs published as a preprint on arXiv, Adlam and Rovlli claim that RQM can be viewed as a time-symmetric theory if all events that have ever occurred and will ever occur are considered as having been created at-once. They also claim that this at-once view (AOV) is necessary to avoid a conflict between RQM and special relativity \parencite{Rovelli2022InformationPhysical}. Now, while the sections of the paper talking about time-symmetry and the AOV were removed in the later published version \parencite{Rovelli2023InformationPhysical}, the introduced postulate of CPLs remained the same with all its implications. Thus, it is still relevant to explore the metaphysical consequences of this new postulate.\footnote{To emphasize that the basics are the same in both versions, I will always quote from the later published version \parencite{Rovelli2023InformationPhysical} when a passage is exactly or almost exactly the same in both versions.}

 While for some people viewing all events as having been created at once is plausible, the AOV is certainly controversial and difficult to reconcile with our manifest image of time. If one does not think, for example, that the arrow of time can be reduced to the arrow of increasing entropy, then it is challenging to understand how the arrow of time enters the world in an AOV. Additionally, if all events are created simultaneously, the (physical) future is already fixed. Thus, the AOV is incompatible with libertarian accounts of free will.\footnote{Importantly, usual deterministic theories like Newtonian mechanics do not generally conflict with libertarian accounts of free will. To create such a conflict, one needs two additional assumptions: First, the deterministic laws and initial conditions are complete, i.e., nothing else can influence the future. And second, the initial conditions are specified with infinite precision. (For reasons why it may make more sense to consider initial conditions as specified with finite precision, see \textcite{Gisin2019PhysicsDeterminism}.)
Normally, therefore, questions of free will can be safely ignored in the development of a physical theory. However, the situation is quite different with the determinism required for the AOV. If the set of all quantum events is already completely determined and one assumes that decisions have physical consequences, all decisions are also completely determined.
Additionally, one should bear in mind that the assumption that we can freely decide what to measure is used as a justification for the assumption of statistical independence  \parencite{Brown1993GhostAtom, Gisin2014QuantumChance, Zeilinger2010DancePhotons} in the derivation of Bell's theorem \parencite{Bell1981BertlmannSocks}. Without this assumption, I can see no reason to reject superdeterminism, a local hidden variable theory of quantum mechanics \parencite{Hossenfelder2020SuperdeterminismGuide, Palmer2020RethinkingSuperdeterminism}. Thus, the AOV creates a justification problem for RQM itself.}
However, these two points only serve to clarify the motivation for the discussion that follows. The sole argumentative aim of this paper is to show that, despite claims to the contrary, an AOV is not necessary for RQM.\\

The structure is as follows: Section \ref{sec: conflict} explains the supposed tension between RQM and relativity. Referring to a paper by Esfeld and Gisin on Ghirardi-Rimini-Weber (GRW) flash theory \parencite{Bell2004SpeakableUnspeakable, Tumulka2006RelativisticVersion, Tumulka2009PointProcesses}, Adlam and Rovelli assume that space-like separated measurement events on an Einstein-Podolsky-Rosen (EPR) pair would lead to a conflict with the relativity of simultaneity and RQM. Section \ref{sec: applicability} analyses whether Esfeld and Gisin's argument can be applied to RQM. Finally, Section \ref{sec:conclusion} offers a brief summary of the results.

\section{GRW Flash Theory or Why the At-Once View Might Be Necessary}\label{sec: conflict}

This section offers a discussion of why the temporal generation of events in RQM might violate relativity, as Adlam and Rovelli claim in the arXiv version of their paper introducing CPLs:

\begin{quote}
    [I]f we want to maintain relativistic covariance then we cannot think of the set of events as being generated in some particular temporal order. [\dots] Thus it seems that RQM is most compatible with a metaphysical picture in which where [sic] the laws of nature apply atemporally to the whole of history, fixing the entire distribution of quantum events all at once. \parencite[13]{Rovelli2022InformationPhysical}
\end{quote}

Adlam and Rovelli do not provide an argument for this statement but refer to a paper by Esfeld and Gisin, which they argue demonstrates that the GRW flash theory is relativistically covariant only if one considers an entire distribution of flashes. Since the ontology of GRW flash theory consists of flashes which, similarly to the events in RQM, occur at every \say{measurement}, it seems initially plausible that Esfeld and Gisin’s argument can also be applied to RQM. The following is therefore a brief summary of the argument put forward by them:\\

Suppose two particles, $S_1$ and $S_2$ are in a Bell state. If two observers, Alice and Bob, each take one of the particles, and move away from each other, they can make two measurements on the particles at space-like separation. If Alice measures the spin of her particle $S_1$, then a flash $f_A$ with result $a$ is produced in GRW flash theory. If Bob measures the spin of $S_2$, then this produces a flash $f_B$ with result $b$.\\

If $f_B$ occurred before $f_A$, then result $a$ depends on $b$ (and Bob’s measurement settings). If $S_1$ and $S_2$ are measured in the same basis, then result $b$ completely determines result $a$. However, if $f_A$ occurs before $f_B$, then result $a$ remains undetermined before Alice’s measurement and is independent of result $b$. Therefore, if an objective fact exists regarding whether $a$ depends on $b$ or is already determined, then an objective fact also exists regarding whether $f_A$ or $f_B$ occurred first. This objective fact about the temporal order of the space-like separated flashes, however, requires a preferred foliation of four-dimensional space-time into three-dimensional hyperplanes, thus violating the relativity of simultaneity. Esfeld and Gisin write the following:
\begin{quote}
    [I]t is not possible to conceive the coming into being of the flashes in a Lorentz-invariant manner. The reason is that the occurrence of some flashes depends on where in space-time other flashes occur: in one frame, Alice’s outcome flash is independent of the flashes that constitute Bob’s setting and outcome; in another frame, Alice’s outcome flash depends on (or is influenced by) the flashes that constitute Bob’s setting and outcome. \parencite[9]{Gisin2014GRWFlash}
\end{quote}

If the flashes do not occur one after another but were all generated at once, then, at the time of Alice’s measurement, it need not be a fact whether $f_B$ occurred before or after $f_A$ since the events were already fixed anyway. Thus, in order not to create a contradiction with relativity, all flashes must be created at-once.\\

The success of Esfeld and Gisin’s argumentation is for GRW flash theory is not relevant here. Instead, the following section explores whether the argument can be applied to RQM.

\section{Applicability to Relational Quantum Mechanics}\label{sec: applicability}

Having presented Esfeld and Gisin’s argument, we can consider whether it indicates that all events in RQM must be created at once to avoid producing a contradiction to the theory of special relativity.

Assume that particles $S_1$ and $S_2$ are in the Bell state $\ket{\Phi^+}=\frac{1}{\sqrt{2}}(\ket{00}+\ket{11})$  and Alice and Bob always measure their particles’ spin in the same basis. When Alice and Bob compare their results, the postulate of CPLs ensures that when Bob asks Alice about her measurement, Alice tells him the result she actually measured. The postulate of internally consistent histories ensures that the answer he hears corresponds to her measurement result. Thus, when Alice and Bob compare their results, they must have obtained the same results in RQM for their measurements in the space-like separation. If Alice’s measurement has already occurred, then Bob can only obtain one possible result. If Alice’s measurement has not yet occurred, then both measurement results are still possible for Bob. Thus, if there is an objective fact about whether one or two measurement results remain possible for Bob, then there is also an objective fact about whether Alice or Bob’s measurement occurred first.
\\

So is there an objective fact about how many measurement results remain possible for Bob? The answer depends on how objectively one views the events themselves in RQM. If a definite fact has objectively occurred in Alice’s measurement of $S_1$, then there is also an objective fact about how many different measurement outcomes remain possible for Bob. Adlam and Rovelli indicate that adding the postulate of CPLs makes events objective and observer-independent. As they write,

\begin{quote}
  [W]ith the addition of CPL, it no longer seems possible to insist that everything is relational---or at least, it is no longer \textit{necessary} to do so---because this postulate implies that the information stored in Alice’s physical variables about the variable $V$ of the system $S$ is accessible in principle to any observer who measures her in the right basis [\dots]. This suggests that the set of \say{quantum events} should be regarded as absolute, observer-independent features of reality in RQM, although quantum states remain purely relational. \parencite[11]{Rovelli2023InformationPhysical}
\end{quote}

Is it necessary to consider the set of events as absolute? Depending on whether one considers systems in RQM to be fundamental, it might not even be possible to view them as such, as one can see by asking what constitutes an \say{event} in RQM in the first place. In RQM, the measurement result—the value that is actualized in the measurement—must initially be valid relative only to the interacting systems. Adlam and Rovelli acknowledge this:

\begin{quote}
  [A]lthough the \textit{event} is an absolute, observer-independent fact, it is still correct to say that the \textit{value} $v$ is relativized to Alice. This is because at this stage Alice is the only observer who has this information about $S$, although other observers could later come to have the same information by interacting appropriately with either Alice or $S$. \parencite[11]{Rovelli2023InformationPhysical}
\end{quote}

Concurrently, however, Adlam and Rovelli do not want to consider systems in RQM to be fundamental:

\begin{quote}
  [A] \say{system} can simply be identified with a set of quantum events that are related to one another in certain lawlike ways, as captured by the formalism of quantum mechanics. [\dots] So, in RQM with CPL, the notion of a system is not necessarily fundamental but rather is used as an interpretative tool to help us make sense of the set of quantum events. \parencite[12]{Rovelli2023InformationPhysical}
\end{quote}

Events, therefore, cannot be defined with reference to systems. If one does not want to consider \say{event} to be a primitive notion, then the only possibility seems to be to characterize events as the actualization of a value of a physical variable. However, actualized values are not absolute, as seen previously. Thus, unless one wants to regard systems in RQM as fundamental or believe the notion of an \say{event} to be primitive, it does not seem possible to consider the set of events as absolute and observer-independent.
But even if one ignores the question of what an event is, it is at least unnecessary to consider the set of events as absolute.\\

Adlam and Rovelli believe that the CPLs make the set of events objective. The CPLs, however, only require that when Bob asks Alice for the result of her measurement on system $S$, she tells him the result she measured. This does not mean that Alice’s measurement must immediately be an objective fact for all systems. The CPLs are equally consistent, with the measurement event (with a definite result) occurring relative only to Alice and $S$. If Bob subsequently interacts with Alice, then he can obtain information about the interaction between Alice and $S$---crucial, as Adlam and Rovelli rightly point out, for avoiding solipsism. In this sense, Alice’s measurement result also becomes a fact relative to Bob. However, this happens only at the moment he interacts with Alice; a fact about Alice’s measurement result for Bob does not need to exist before that. For other systems that do not interact with Alice or $S$, there need not be any facts about the measurement.\\

Consequently, in the above entanglement scenario, it is not that a definite fact is objectively actualized in Alice’s measurement of $S_1$; rather, a fact about the measurement is actualized (apart from Alice and $S_1$) only for Bob and only at the moment he interacts with Alice. So there need not be an objective fact about how many measurement outcomes are possible for Bob when he interacts with $S_2$. From Alice’s perspective, it may be that only one outcome is possible for Bob. However, this is not the case for Bob. The no-signalling theorems \parencite{Weber1980GeneralArgument} show that his probability of obtaining a particular measurement result does not change because of Alice’ measurement of the spin of $S_1$. Thus, if events are not considered objective and observer-independent, then the argument developed for GRW flash theory cannot be applied to RQM. Esfeld and Gisin acknowledge that the conflict between quantum mechanics and relativity can be avoided if such an objective view of measurements is abandoned.

\begin{quote}
    [O]ne may envisage to maintain that what there is in nature depends on the choice of a hypersurface—so that different facts exist in nature relative to the choice of a particular foliation of space-time. However, if one is willing to endorse such a relativism, any of the known proposals for a primitive ontology of QM can then easily be made relativistic. \parencite[11]{Gisin2014GRWFlash}
\end{quote}

There are several concepts for how different facts can exist relative to different foliations. See, for example, \textcite{Fleming1996JustHow, Fine2006RealityTense, Myrvold2002PeacefulCoexistence, Myrvold2003RelativisticQuantum, Myrvold2016LessonsBell, Myrvold2019OntologyRelativistic, Myrvold2021RelativisticConstraints}. However, what is special about RQM is that this relativization of facts to systems comes much more naturally than the relativization to foliations in other interpretations of quantum mechanics.\\

\begin{figure}[h]
    \centering
    \includegraphics{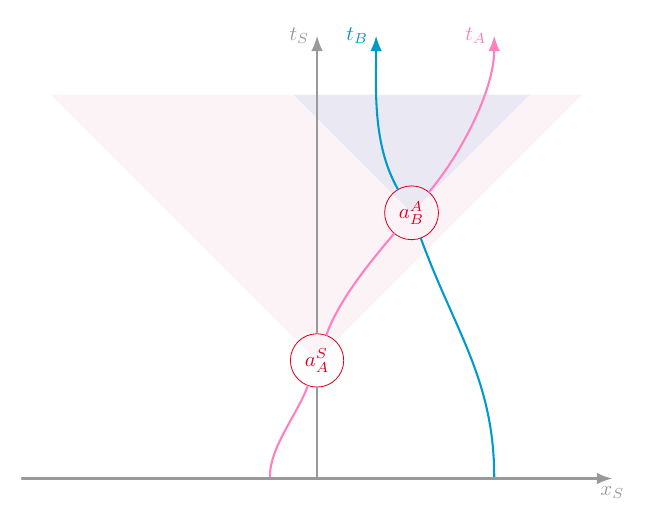}
    \caption{Space-time diagram of a situation in which the CPLs apply: Since Alice must be on a time-like curve, the events $a_A^S$ in which Alice performs a measurement on system $S$ with outcome $a$ and the event $a_B^A$ in which Bob asks Alice for her information about $S$ must be time-like separated. The CPLs, therefore, cannot require a temporal order between space-like separated events. They do not apply to them. Thus, the CPLs don’t create a conflict between RQM and special relativity.}
    \label{fig:event-net CPL}
\end{figure}

In general, it is implausible that the CPLs could create a conflict between RQM and relativity and thus necessitate the AOV. In a scenario in which the CPLs apply, Alice must be on a time-like curve between her measurement of $S$ and her interaction with Bob (see Figure \ref{fig:event-net CPL}). Thus, the CPLs can only relate two time-like separated events, which is unproblematic in special relativity theory. Consequently, it is incomprehensible to me how adding CPLs to RQM could compel an AOV.\\

A possible problem arises, however. The CPLs require that the two measurement results be correlated when Alice and Bob compare their results. If Alice and Bob do not compare their results, then nothing ensures that the two measurement results are correlated. So if a later comparison of the measurement results ensures that they are correlated, then is this not retrocausality, which presupposes a form of AOV?
(E. Adlam, personal communication, September 2022)
\newline

\begin{figure}[h]
    \centering
    \includegraphics{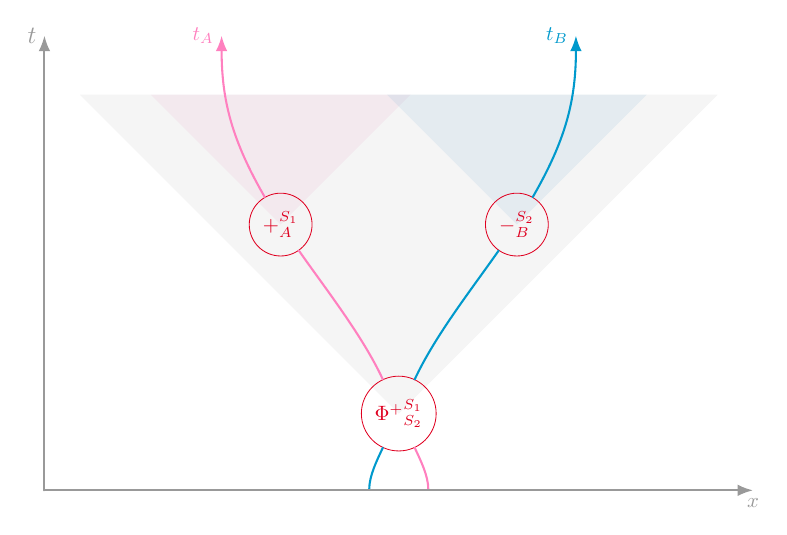}
    \caption{Space-time diagram of an entanglement scenario in which Alice and Bob obtain incompatible results. Although they should have the same result due to the entanglement of their particles in the $\Phi^+$ state, Bob measures spin \protect\say{up} in the $x$-direction ($+$) and Alice spin \protect\say{down} ($-$). In RQM, this problem does not arise because there is no perspective from which measurement results are compatible or incompatible, as long as they are not compared with each other.}
    \label{fig:event-net Entanglement gone wrong}
\end{figure}

One might ask what, in the entanglement scenario above, would prevent Alice from measuring $+$ and Bob from measuring $-$ in space-like separation   such that their results are not correlated according to the laws of quantum mechanics (see Figure \ref{fig:event-net Entanglement gone wrong}).\\

The problem is solved by asking from which perspective the measurement results are correlated or not. Before Alice and Bob compare their results, there is no fact for any system about whether the events are correlated. Stating that Alice measures $+$ and Bob measures $-$ in space-like separation and drawing the space-time diagram in Figure \ref{fig:event-net Entanglement gone wrong} presupposes a \say{God’s eye view}, which---at least traditionally---does not exist in RQM \parencite{Rovelli2022RelationalQuantum, sep-qm-relational, Rovelli1996RelationalQuantum}. Only in the interaction between Alice and Bob does a fact emerge regarding whether the measurement results are correlated. There is no retrocausal change from uncorrelated to correlated when Alice and Bob decide to compare their results, since there was no fact about the correlation beforehand.\\

Depending on their world lines, either Bob or Alice’s measurement will have happened first for them. There is no objective fact about which measurement occurred first. From one perspective, Bob’s measurement might have determined Alice’s, and from another, vice versa. Nothing in RQM requires an objective fact about this to exist. Thus, there can be temporal development of the set of events, albeit a complicated one (different for each system), as in any account of relativistic temporal development \parencite{Ellis2008FlowTime, Gisin2021RelativityIndeterminacy, Fleming1996JustHow}.

\section{Conclusion}\label{sec:conclusion}

The purpose of this paper was to address whether a temporal development of events is possible in RQM or whether all events must have been created at once. If an objective fact exists regarding whether the occurrence of Alice’s measurement event depends on Bob’s space-like separated measurement event, then an objective fact must also exist about which event occurred first, which leads to a contradiction with the relativity of simultaneity.

The present paper demonstrated that there need not be such an objective fact about the dependence of Alice’s measurement event on Bob’s if one does not regard the set of events in RQM as absolute. Moreover, it could be shown that even with the postulate of CPLs, one does not need to consider the set of events in RQM as absolute. Thus, contrary to Adlam and Rovelli’s statement, the AOV is unnecessary, and the events of RQM can be generated temporally.

\section*{Acknowledgements}
I sincerely thank Lukas Schmitt and Emily Adlam for the extensive discussions on this topic and their many helpful suggestions.\\

\newpage


\printbibliography
\newpage
\end{document}